\documentclass[aps, prl,preprint,superscriptaddress,showpacs,amsmath,amssymb]{revtex4-1}
\usepackage[dvipdfm]{graphicx}% Include figure files
\usepackage{dcolumn}% Align table columns on decimal point
\usepackage{bm}% bold math
\usepackage{multirow}% Author names in capital letters:

\begin{document}

%% ------------------------------------------------------------------------ %%
%
%  TITLE
%
%% ------------------------------------------------------------------------ %%

\title{Hardening and termination of long-duration gamma rays detected prior to lightning}
\author{H. Tsuchiya}
\affiliation{Japan Atomic Energy Agency, Tokai-mura, Naka, Ibaraki 319-1195, Japan}
\affiliation{High-energy Astrophysics Laboratory, Riken, 2-1, Hirosawa, Wako, 
Saitama 351-0198, Japan}
\author{T. Enoto}
\affiliation{High-energy Astrophysics Laboratory, Riken, 2-1, Hirosawa, Wako, 
Saitama 351-0198, Japan}
\affiliation{Goddard Space Flight Center, NASA, Greenbelt, Maryland, 20771, USA}
\author{K. Iwata}
\affiliation{Shibaura Instirute of techmology, Minuma, Saitama, Saitama 337-8570, Japan}
\author{S. Yamada}
\affiliation{High-energy Astrophysics Laboratory, Riken, 2-1, Hirosawa, Wako, Saitama 351-0198, Japan}
\author{T. Yuasa}
\affiliation{Department of High Energy Astrophysics,
Institute of Space and Astronautical Science, JAXA, 3-1-1,
Chuo-ku, Sagamihara, Kanagawa 252-5210, Japan}
\author{T. Kitaguchi}
\affiliation{High-energy Astrophysics Laboratory, Riken, 2-1, Hirosawa, Wako, Saitama 351-0198, Japan}
\author{M. Kawaharada}
\affiliation{Department of High Energy Astrophysics,
Institute of Space and Astronautical Science, JAXA, 3-1-1,
Chuo-ku, Sagamihara, Kanagawa 252-5210, Japan}
\author{K. Nakazawa}
\affiliation{Department of Physics, University of Tokyo, 7-3-1, Hongo, 
Bunkyo-ku, Tokyo 113-0033, Japan}
\author{M. Kokubun}
\affiliation{Department of High Energy Astrophysics,
Institute of Space and Astronautical Science, JAXA, 3-1-1,
Chuo-ku, Sagamihara, Kanagawa 252-5210, Japan}
\author{H. Kato}
\affiliation{High-energy Astrophysics Laboratory, Riken, 2-1, Hirosawa, Wako, 
Saitama 351-0198, Japan}
\author{M. Okano}
\affiliation{High-energy Astrophysics Laboratory, Riken, 2-1, Hirosawa, Wako, 
Saitama 351-0198, Japan}
\author{T. Tamagawa}
\affiliation{High-energy Astrophysics Laboratory, Riken, 2-1, Hirosawa, Wako, 
Saitama 351-0198, Japan}
\author{K. Makishima}
\affiliation{Department of Physics, University of Tokyo, 7-3-1, Hongo, 
Bunkyo-ku, Tokyo 113-0033, Japan}

\date{\today}% It is always \today, today,
%  ABSTRACT
%
%% ------------------------------------------------------------------------ %%

% >> Do NOT include any \begin...\end commands within
% >> the body of the abstract.

\begin{abstract}
We report the first observation of 3$-$30 MeV prolonged gamma-ray emission that was abruptly terminated by lightning. 
The gamma-ray detection was made during winter thunderstorms on December 30, 2010 by the Gamma-Ray Observation of Winter THunderclouds (GROWTH)
experiment carried out in a coastal area along the Sea of Japan.
The gamma-ray flux lasted for less than 3 min, continuously hardening closer to the lightning occurrence. 
The hardening at energies of 3$-$10 MeV energies was most prominent.  The gamma-ray flux 
abruptly ceased less than 800 ms before the lightning flash that occurred over 5 km away from the experimental 
site.  In addition, we observed a clear difference in the duration of the 3$-$10 MeV gamma rays and those $>$10 MeV, 
suggesting that the area of $>$10 MeV gamma-ray emission is considerably smaller than that of the lower-energy gamma rays.  
This work may give a manifestation that a local region emitting prolonged gamma rays connects with a distant region to 
initiate lightning. 
%This work implies that electrons emitting the observed prolonged gamma rays are the same as those causing the lightning initiation.
\end{abstract}
\maketitle
%% ------------------------------------------------------------------------ %%
%
%  BEGIN ARTICLE
%
%% ------------------------------------------------------------------------ %%

% The body of the article must start with a \begin{article} command
%
% \end{article} must follow the references section, before the figures
%  and tables.

\section{Introduction}
Prolonged gamma rays emitted from thunderclouds were observed during the 1980s and 1990s 
by detectors onboard an airplane and a balloon~\citep{MP_1985,Eack_1996}.
High-mountain experiments~\citep{EAS_2000, Alex_2002,MtFuji_2009,norikura_2009,Chili_2010,Chili_2011,Tsuchiya_tibet_2012} 
and sea-level measurements~\citep{MONJU_2002,growth_2007,MONJU_2011,growth_2011} also observed thundercloud-related
gamma rays that lasted for a few seconds to $\sim$10 min, and even 40 min
on rare occasions. 
%In addition, it was also found that the maximum energy of these gamma rays reached 10 MeV or higher.
%Thus, as originally predicted by \citet{Wilson}, it has been established that 
%electrons are continuously accelerated to relativistic energies by quasistatic electric fields in thunderclouds,
%producing high-energy emissions via bremsstrahlung. 
However, we do not fully understand the causes for 
the differences in duration and the temporal variations of the individual thundercloud-related gamma-ray events.
Better understanding these parameters would provide valuable information
on the charging mechanism of thunderclouds as well as the electric-field acceleration that
occurs in natural accelerators.

%Generally, prolonged gamma rays are not produced by lightning discharges, but short-duration
Unlike prolonged gamma rays, short-duration ones are related to natural lightning~\citep{Moore_Obs2001,Dwyer_Obs2005,Yoshida_Obs2008} 
and artificial lightning~\citep{Dwyer_Obs2004}. These lightning-related gamma rays last for a millisecond or less.
%, and typically have energiesof a few hundred keV or a few MeV on rare occasions. 
Although there are distinct differences between 
thundercloud- and lightning-related gamma rays, especially in duration, they are considered to have a common
viable mechanism based on the relativistic runaway electron avalanche models~\citep{Gurevich_RREA_Cal1992,Dwyer_MC2003,Babich_MC2004}. 
%The model was further developed by \citet{Dwyer_MC2003} and \citet{Babich_MC2004}, 
These models explain the generation of large amount of nonthermal photons by ambient electrons as follows.
Energetic electrons, possibly seeded by {\it e.g.} cosmic rays, will run away %from the electric field 
when the gained energies from field are
sufficiently high to exceed the ionization loss in air.
Then, %increasing in the number of electrons, 
those electrons will produce a detectable flux of bremsstrahlung photons. 
Presently,  a large number of electrons produced by 
electron avalanches may cause the lightning initiation %~\citep{Babich_Cal2012}.
through an enhancement of the electric fields and the conductivity of the atmosphere~\citep{Babich_Cal2012}. 
%Those current models have not yet well explained whether thundercloud-related gamma rays connect with lightning-related ones. 

Contrary to the general characteristics of prolonged gamma rays, 
a few observations have provided a possible association with lightning occurrence. 
Airplane-~\citep{MP_1985} and a balloon-based measurements~\citep{Eack_1996} showed that
x-ray flux at energies of 1$-$100 keV suddenly ceased at lightning occurrence. 
In addition, a high-mountain experiment~\citep{Alex_2002} demonstrated that the particle count increased at energies 
of $>$30 MeV then quickly returned to the background level when lightning occurred.
These events %accompanied by sudden count increase terminations prior to lightning,  
suggest that electrons are continuously accelerated to relativistic energies until just before the lightning strike. 
%Thus, there may be a possible connection between the prolonged gamma rays and lightning. 
However, it is still unclear whether runaway electrons producing long-duration gamma rays
are the same as those related to lightning initiation. 
Here we present the results of a measurement of one prolonged gamma-ray event with the finest time resolution of 0.1 ms
and compare it with previous observations of long-duration gamma-ray events.
\section{Experiment}
The Gamma-Ray Observation of Winter THunderclouds (GROWTH) experiment has been 
operating successfully at the Kashiwazaki-Kariwa nuclear power plant since December 22, 2006. 
The experimental site is located in the coastal area of the Sea of Japan where winter lightning is common.
The data acquisition system, daq1  
was installed on the roof of a building on December 2, 2010. It is 
arranged in a north-east to south-west direction 780 m apart from the original system, daq0. 
Detailed information on daq0 can be found in \citet{growth_2007,growth_2011}. Thus,
we provide an outline of the daq1.

%Daq0 has two detectors, detector A and detector B~\citep{growth_2007,growth_2011}, for continuous
%measurements of energetic radiations.  They consist of inorganic NaI, CsI, and BGO scintillators.
%Since we did not use detector A in this study (due to hardware trouble), we briefly describe detector B.
%It consists of $\phi$7.62-cm spherical NaI and CsI scintillation counters. The former
%observes 0.04$-$11 MeV, while the latter observes 0.03$-$80 MeV. 
%Each counter accumulates pulse-height spectra in 1024 channels every 6 sec and records broadband pulse counts every second. 
%In addition to the radiation counters, daq0 has an electric-field mill (EFM-100). 
%The output of this mill is recorded as analogue to digital converted (ADC) values every second.  
%The sampling time for the radiation detectors and electric-field mill are synchronized with 1-second pulses of the Global Positioning System (GPS).
 
Daq1 has a cylindrical $\phi$7.62$\times$7.62 $\mathrm{cm^2}$ NaI scintillator equipped with a photomultiplier (HAMAMATSU R1306).
Above the NaI counter, a thin plastic scintillator with a thickness of 0.5 cm and an area 30$\times$30 $\mathrm{cm^2}$
is placed. The plastic scintillator is wrapped in thin aluminum foil with a thickness of 15 $\mu$m and then covered with 
a black 100 $\mu$m-thick sheet. The plastic scintillator is connected to a photomultiplier (HAMAMATSU R1306) 
by a light guide. It is utilized to reject charged-particle background (mostly muons) by the anti-coincidence method. 
%The anti-coincidence means that an event in the NaI scintillator is judged as a charged particle
%if the plastic scintillator simultaneously (in 1 $\mu$s) give a signal. Thus, using signals of the NaI and plastic scintillators
%both in anti-coincidence, we can effectively select photons. 

The signals of the NaI and plastic scintillators are fed to a 12-bit ADC chip (AD7862-10) incorporated in a self-triggering 
electronics system and recorded as ADC values corresponding to the energy deposits in the individual scintillators. 
The arrival time of each event is determined by the GPS synchronized with 
a 10 kHz frequency.  Daq1 observes an energy 
range of 0.04$-$30 MeV with a time resolution of 0.1 ms. Because daq0 also employs a GPS to determine the recording time, 
the time of the two GROWTH systems coincide. Moreover, daq1 has an optical sensor with
a Si photodiode (HAMAMATSU S1226-8BK) for wavelengths 
of 320$-$1000 nm, having a peak of 720 nm. The output signal of the photodiode is fed to a peak-hold ADC, and
the peak value per second is recorded as the ADC value.

In addition to the two systems of daq0 and daq1, nine monitoring posts (MPs) are operated by the Tokyo electric power company 
to monitor environmental radiation doses (circles in Fig.1 of \citep{growth_2011}). Each MP consists of a cylindrical $\phi$5.1$\times$
5.1 $\mathrm{cm^2}$ NaI scintillation counter and an ion chamber. The former covers the energy range 50 keV$-$3 MeV, and
the latter measures for those $>$50 keV. The time resolution of the two detectors is 30 s. MPs are widely distributed in the premises
at intervals of 300$-$400 m.

\section{Results}\label{sec:res}
\begin{figure}
 \noindent\includegraphics[scale=0.6]{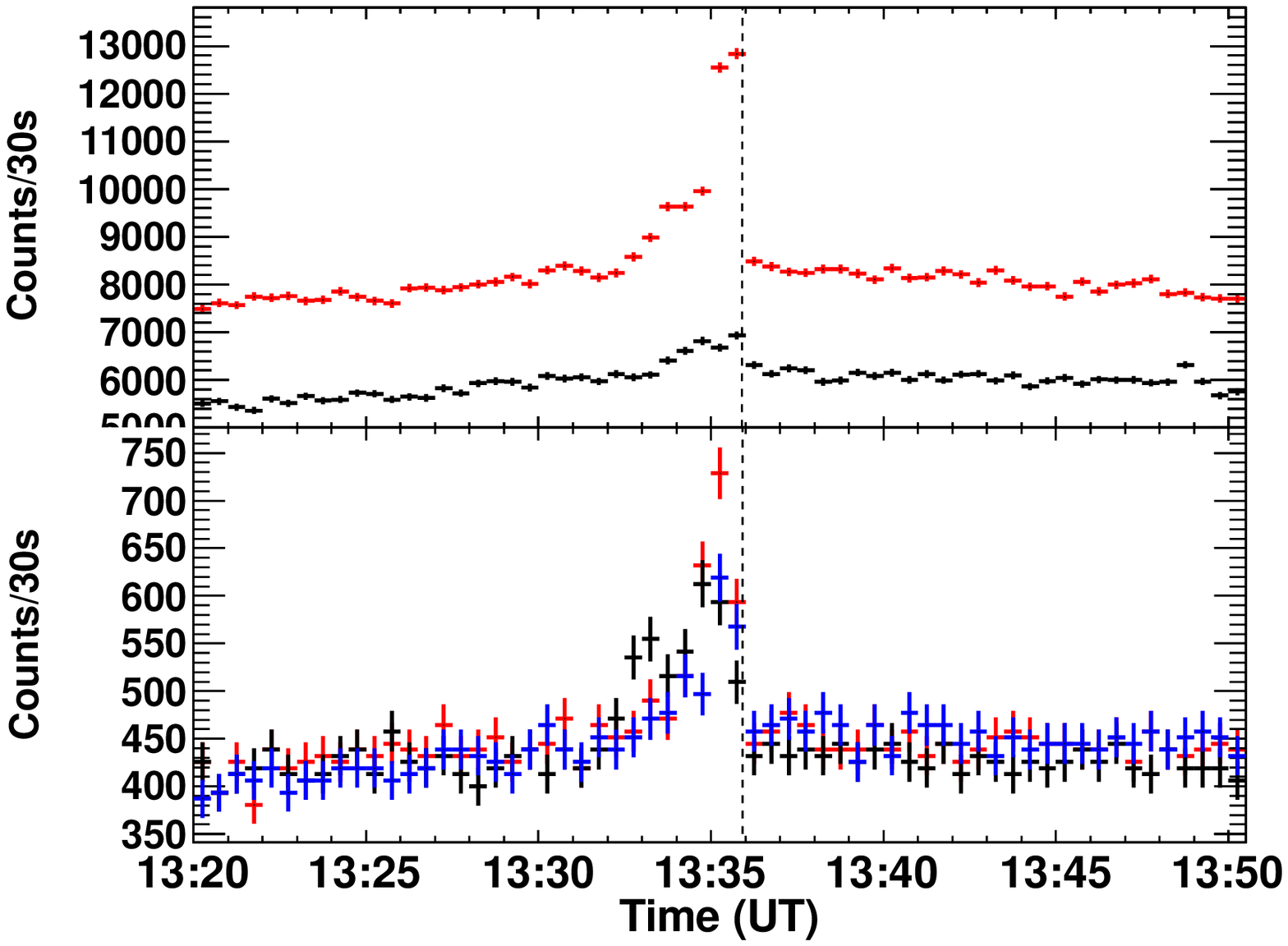}
 \caption{Count histories per 30 s of daq0, daq1, and MPs 7$-$9, obtained during 13:20$-$13:50 UT on December 30, 2010. 
(Top) The count rates for $>$40 keV from the NaI scintillators of daq0 (black) and daq1 (red) without anticoincidence. 
(Bottom) The 0.05$-$3 MeV count rates from the NaI scintillators of MPs 9 (black), 8 (red), and 7 (blue).
The vertical error bars indicate the 1$\sigma$ standard deviation. The vertical dashed line shows 
the occurrence time of the lightning flash (13:35:55 UT)}
 \label{fig:30scounts}
%\noindent\includegraphics[scale=0.7]{Figures/Kashiwazaki_daq01_MPs_eng.eps}
% \caption{A bird eye's view of the Kashiwazaki-Kariwa nuclear power plant. Original image
% is taken from Google Map.}
% \label{fig:map}
 \end{figure}

Figure~\ref{fig:30scounts} shows 30-s count rates of the NaI scintillators obtained from daq0, daq1, and MPs 7$-$9 
(Figure~\ref{fig:image} here, Fig. 1 of \citep{growth_2011}) 
obtained over the period of 13:20$-$13:50 UT %(22:20$-$22:50 Japanese Standard Time) 
on December 30, 2010. 
For comparison, the count rates of daq1 (upper crosses in the top panel) show results from the NaI scintillator without anticoincidence.
%that clearly provided enhanced counts during 13:32$-$13:36 UT. 
The background level of daq1 was $\sim$1.4 times higher than that of daq0 (lower crosses in the top panel)
because the effective volume of daq1 %whose detector is a cylindrical type scintillator,  
was 1.5 times larger than that of daq0. %whose detector is a spherical type scintillator. 
The counts increased from 13:32 UT until 13:36 UT, over a time scale which is typical for  
prolonged thundercloud gamma-ray emission~\citep{MONJU_2002,growth_2007,MONJU_2011,growth_2011}.
However, at 13:35:55 UT, the counts suddenly dropped  to the background level.
Such a sudden termination of the gamma-ray emission was never 
observed during previous long-duration GROWTH events~\citep{growth_2007,growth_2011}.

From the optical sensor and electric field mill data it can be seen that lightning occurred 
at 13:35:55 UT, which coincided with the gamma-ray termination. 
The lightning  event is assumed to have occurred at a distance of $>$5 km from the experimental site.
No lightning within 5 km of the site was recorded over the period between 13:05 and 14:05 UT
by the Japan Lightning Detection Network system (operated by Franklin Japan Co. Ltd.),  
which has a lightning detection efficiency of $>$90\%. 
One intracloud discharge occurred 1.7 km south-east of 
the site, but it was at 13:39:25 UT, much later than the gamma-ray termination.
Such a low occurrence of lightning in winter is relatively common
in this region~\citep{Michimoto_Obs1993}. 

The most intriguing aspect of this gamma-ray event is how its sudden termination is related (or
unrelated) to the distant lightning. This can be investigated by estimating the horizontal 
spread of the gamma-ray emitting region.
Using the 30 s count data at energies between 50 keV and 3 MeV, 
%With the 30-s count increases at energies between 50 keV and 3 MeV, we roughly estimated the horizontal spread.
we first calculated the count increases (in percent) 
and the statistical significance of data from all the detectors measured between 13:34 and 13:36 UT. 
%Here, the background level for each detector was evaluated as a quadratic curve derived by a $\chi^2$ fitting. 
Then using the count increases and positions of each detector, an ellipse-image fitting 
was performed according to a method that was originally developed for higher-energy 
gamma-ray observations~\citep{Hillas_image}. The fitting evaluates the major and minor axes of an ellipse. 
In this work, the major and minor axes represent the root mean square 
spread of the radiation in directions along and perpendicular to the axis, respectively, 
connecting the center of the ellipse and the detector position 
with the maximum count increase at each time interval. 
Since small amplitude signals measured by distant detectors may distort the estimated image, 
we did not use individual count increases with statistical significance $<$2$\sigma$.
In practice, MPs 1$-$6 showed no count increases with statistical significance $>$2$\sigma$ during 13:34$-$13:36 UT.
Table~\ref{tab:sigma} lists the numerical values used in the evaluation. 
%The long duration together with the lack of count increases in MPs 1$-$6 excludes the present 
%burst being due to electric noise caused by thunderstorms.

Figure~\ref{fig:image} shows the count-intensity distributions along with 
the radiation spread determined by the above method. The diameters of the 
circles correspond to the individual count increase (in percent). 
Thus, larger diameters correspond to higher count increases. 
%As shown, the determined ellipses have a relatively large eccentricity 0.7$-$0.9. This might show that 
%an energetic radiation source area in the thunderclouds change its form due to thundercloud charging.
Though probably affected by using a small number of detectors, 
the horizontal extent %of the radiation incident on the ground 
was at most $\sim$800 m, which is consistent with 
other observations~\cite{MONJU_2002,MONJU_2011,growth_2011}. 
In addition, the illuminated region appeared to 
move slowly  from southwest to northeast and approach daq1 and MP8. 
Such movement of  a gamma-ray emitting region
was previously reported without performing an ellipse fitting.~\cite{MONJU_2002,MONJU_2011,growth_2011}.
As seen from Fig.~\ref{fig:image} and Table~\ref{tab:sigma}, daq1 was the closest detector to 
the gamma-ray source in the thunderclouds between 13:35:30 and the time that the lightning occurred.
%Here, the count increases of daq0 for each time interval have 5$-$10$\sigma$ statistical significance, while
%those of MPs 5 and 6, close to daq0, did not provide such high statistical significance.
%One of the main reasons is that the effective volume of daq0 is larger than that of MPs 5 or 6 by a factor of 2.3.

Figure~\ref{fig:growth_MPs} shows data from the NaI scintillator with anticoincdence measured by daq1, 
and we can conclude that the observed count increases were attributable to gamma rays.
% shows the 6-sec count profiles of daq0 and daq1 in two higher energy bands.
%Sampling the data of the NaI scintillator with anticoincdedence for daq1,  the observed count increases 
%can be attributed to gamma rays. 
In addition, it is found that the count histories of daq1 exhibited more significant enhancements 
in two energy bands than the data from daq0. Particularly, the count increases in the 3$-$10 MeV band are the most prominent. 
The net count increases of daq0 and daq1 obtained during 13:33:12$-$13:35:55 UT for the 3$-$10 MeV band are $140\pm15$ (9.3$\sigma$)  
and $950\pm30$ (32$\sigma$), respectively. Here, background levels per 6 s for the 3$-$10 MeV energy band were estimated to be
$2.88\pm0.12$ and $2.53\pm0.12$ for daq0 and daq1, respectively, using a constant fitting of the data obtained excluding the above period.
%The count increases for daq1, obtained between 13:34 and the time of the lightning, can be well expressed by an 
%exponential function with a time constant of $\sim$50 s ($\chi^{2}/{\it d.o.f.}=25.5/17$) rather 
%than a linear function ($\chi^{2}/{\it d.o.f.}=56.0/17$). 
Furthermore, count rates for $>$10 MeV from daq1 with anticoincidence (bottom of Fig.~\ref{fig:growth_MPs})  
provided a remarkable net count increase of $57\pm8$ over the period 13:35:19$-$13:35:55 UT above the background level of $2.18\pm0.11$, 
while that of daq0 exhibited a net count increase of 17$\pm$7, which was not statistically significant, during the same time interval,
with a background level of $6.1\pm0.2$.
The statistically significant $>$10 MeV gamma rays were first detected from 
sea-level observations. 
%For comparison, the count increases of daq1 without anticoincidence at $>$10 MeV provided 
%a net count increase of $71\pm11$ ($9.0\pm0.1$) during the same interval. Here, the value in parentheses is the 
%background level of daq1 without anticoincidence at $>$10 MeV.
%These results suggest that the insignificant detection of $>$10 MeV photons by daq0 did not arise from the poor 
%background rejection without its anticoincidence mode of daq0, but probably 
%from the lack of a detectable flux of $>$10 MeV gamma rays arriving at daq0.  

The top panel of Fig.~\ref{fig:daq1_E_vs_t} shows 
time variations of the energies of individual photons recorded by daq1.
Several horizontal stripes can be observed at the 0.3$-$3 MeV band, corresponding to the natural environmental gamma-ray 
lines including 609 keV (${}^{214}$Bi), 1.46 MeV (${}^{40}$K), and 2.62 MeV(${}^{208}$Tl).
%that are used to calibrate the energy scale for daq0 and daq1. 
Clearly, the number of photons with energies of $>$3 MeV 
dramatically increased before the lightning occurrence.
We applied a waiting time function of $r\exp{(rt)}$ to the data of 0.04$-$0.3, 0.3$-$3, 3$-$10, and $>$10MeV 
during 13:33:12$-$13:35:55 UT and derived $r=330\pm2$, $230\pm2$, $9.0\pm0.4$, and $1.4\pm0.3$, respectively. 
Here, $r$ represents the average arrival rate of photons (per s). %(or the inverse of $r$ denotes the average time interval between sequential 
%detections in the same energy band).
For comparison, we calculated $r$ for data outside the burst interval as
$250\pm1$, $170\pm1$,  $1.42\pm0.11$, and $0.40\pm0.05$ for the same four energy bands.
Dividing  by the individual arriving rates of the background levels, the individual rates during the count increases were 
enhanced by a factor of 1.3, 1.4, 6.4, and 3.5, respectively. Consequently, 
the count increases for the two higher energy bands were much greater than those for the two lower-energy 
bands prior to the initiation of lightning, implying that the gamma-ray energy spectrum became harder before 
lightning.

As shown in the bottom panel of Fig.~\ref{fig:daq1_E_vs_t}, 
the $>$3 MeV gamma-ray hardening abruptly ceased 
within 0.1 ms or less, 800 ms before the lightning.
No $>$3 MeV gamma rays were detected by daq1 for $\sim$1 s from the time 800 ms before the lightning.
Unlike the photons with $>$3 MeV, those with $<$3 MeV had no clear gap around the lightning flash.
%In practice, the lightning flash occurred during 13:35:55$-$13:35:56 UT; this is based on
%the output of the optical sensor sampled by a peak-hold ADC at 1 s intervals
%synchronized by GPS. Therefore, the time interval between the gamma-ray termination and the lightning occurrence 
%ranged from 0.8 s to 1.8 s, 
These results suggest that the thundercloud electric fields stopped
accelerating the electrons toward the ground at least 800 ms prior to the lightning flash. 
\section{Discussion}
\citet{MP_1985} and \citet{Alex_2002} have observed similar events to that under discussion, who conducted 
an airplane observation and a high-mountain experiment, respectively. %The former showed that in 2$-$10 s, 
%the 5$-$110 keV x-ray flux increases and returns to the background level $\sim$100 ms prior to the lightning flash, 
%while the latter observed an abrupt termination of $>$30 MeV radiation with a time resolution of 1 s. 
Unlike these two past observations, the current results, employing one photon counting with a time resolution of 0.1 ms, 
demonstrate that 3$-$30 MeV gamma rays clearly terminate 800 ms prior to the lightning flash (Fig.~\ref{fig:daq1_E_vs_t}). 
%In addition, the present results gave no lightning-related emissions within 800 ms.
In actuality, this time scale of 800 ms agrees generally with the total discharge duration, which is a few hundred to 700 ms, 
of intracloud discharges~\citep{lightningbook}. 

In addition, the present event and the two past observations did not 
detect any energetic x-ray radiation associated with lightning, mainly its stepped leaders.
Therefore, the lack of detection of such x-ray energetic radiation may be a common feature, implying 
that the source of short-duration x-ray bursts is either different or far from that of the precursory prolonged emission. Alternatively,
the lightning-related x rays may not have been beamed toward the detectors. Then, these  events also 
imply that a local electric field gradually (a few seconds to a minute) enhances.
\citet{Babich_Cal2012} recently showed via numerical simulations that 
such an electric-field enhancement which can potentially generate lightning can last for a relatively long duration of $\sim$10 s via 
electron avalanches caused by a steady flux of secondary cosmic-ray electrons.
Thus, the present temporal behavior prior to the gamma-ray termination (Fig.~\ref{fig:growth_MPs}) might indicate such an effect.

%In addition, recent high-mountain experiments \cite{Gurevich_Obs2011} have revealed that lighting can occasionally 
%produce $>$40 keV energetic emissions lasting 100$-$600 ms, instead of $<$10-ms duration ones. They indicated that such a longer
%emission, compared with  usual ligntining-related ones, is associated with whole processes 
%of lighting rather than part of the lighting processess (e.g. stepped leaders).
Hence, we may interpret the lack of gamma-ray detection during this time interval in the following manner.
Through a process as proposed by \citet{Babich_Cal2012}, the field strength of the acceleration region was enhanced, 
for a prolonged period, until its strength was high enough 
to produce the observed gamma rays. Then, certain lightning processes, such as stepped 
leaders, emerged from the acceleration region to initiate an intracloud discharge. 
As described in Section 3, % indicated by the observation of the Japan Lightning Detection Network system as mentioned in Section 3, 
the observed lightning occurred more than 5 km away from our detectors. 
%An energetic radiation may have been generated from intracloud lightning having a duration of $\sim$800 ms.
%all the processes associated with the discharge having a duration of 800 ms. 
%The energetic radiation did not reach our detectors because of the large distance ($>$5 km), 
%but certain lightning process, such as stepped leaders, possibly arrived at somewhere in the acceleration 
Thus, as often observed~\citep{Proctor_1991}, 
the stepped leaders would, either horizontally or in an upward direction, propagate over the distance towards another charged region of the thundercloud.
This would be possible because lighting paths can extend over 2$-$8 km (e.g. \citep{Weber_1982}). 
%Because such stepped leaders propagate with an average speed of 
%$\sim10^{5}$ $\mathrm{m/s}$~\citep{ightningbook}, and would reach the acceleration region in 50$-$80 ms
Consequently, a connection between the acceleration region and the other charged region could neutralize the acceleration region 
$\sim$800 ms before the optical flash was observed, resulting in the termination of the electron acceleration. 
Then, the field region might not recover to a high enough level to accelerate electrons to relativistic energies, 
at least until the end of the intracloud lightning event. 

\begin{figure}
\noindent\includegraphics[scale=0.6]{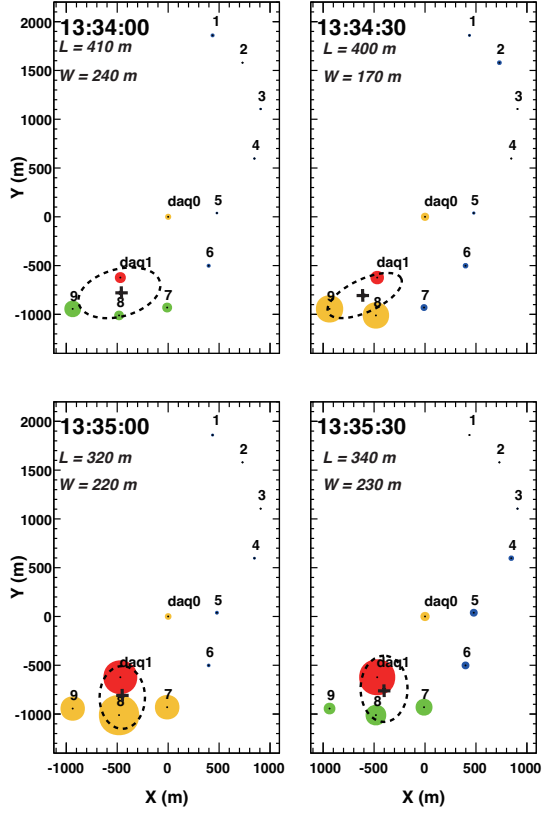}
\caption{Distributions of count enhancements for all detectors in four time intervals.
The values 1$-$9 represent the MP location.
{\it L} and {\it W} in each panel are the 
lengths of the major and minor axes of the ellipses (dashed lines), respectively. The crosses show
the center of the ellipses. Colors of circles indicate statistical significance obtained by individual detectors. Blue, green, 
orange, and red represent those $<$2$\sigma$, 2$-$5$\sigma$, 5$-$10$\sigma$, and $>$10$\sigma$,
respectively.}
\label{fig:image}
 \end{figure}

\begin{figure}
 %\noindent\includegraphics[scale=0.6]{Figures/Fig_daq01_lcs_3MPs_nai_1320-1350_30sec.eps}
\noindent\includegraphics[scale=0.6]{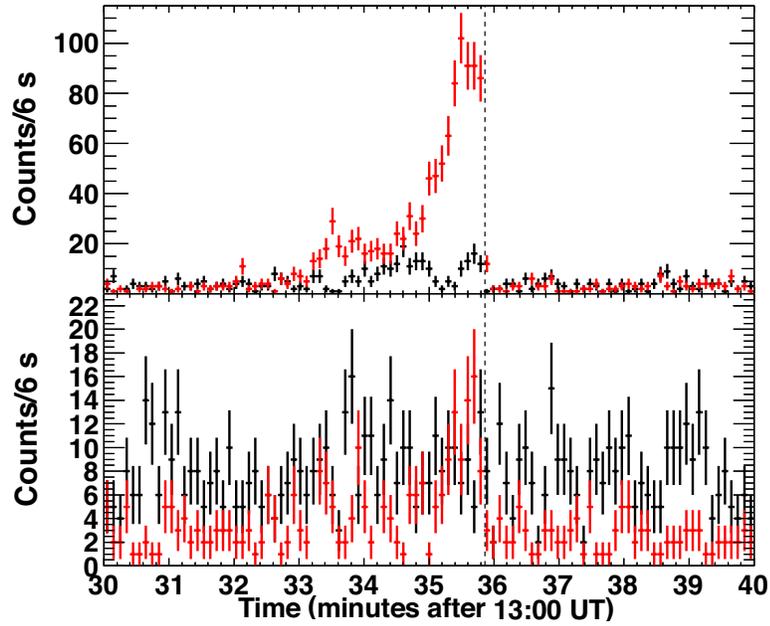}
%\noindent\includegraphics[scale=0.6]{Figures/Fig_daq01_2Ebands_lcs_6sec_1330-1340.eps}
%\noindent\includegraphics[scale=0.5]{Figures/Fig_daq01_ld_MPs_nai_1320-1350_30sec.eps}
%\noindent\includegraphics[scale=0.5]{Figures/Fig_daq01_4Ebands_lcs_30sec_1310-1400.eps}
 \caption{%(Left) Count histories per 30 s of daq0, daq1, and MPs, obtained over 13:20$-$13:50 UT  on 2010 December 30. 
%(a) The $>$40 keV count rates from the NaI scintillators of daq0 (black) and daq1 (red) without anti-coincidence. 
%(b) The 0.05$-$3 MeV count rates from the NaI scintillators of monitoring posts 9 (black), 8 (red), and 7 (blue).
Count histories per 6 s for radiations observed at daq0 (black) and daq1(red), obtained during 13:30$-$13:40 UT.
(Top) Count histories of 3$-$10 MeV of  the NaI scintillators of daq0 and daq1 with anticoincidence mode.
(Bottom) Same as that of the top panel, but for $>$10 MeV.  
Vertical error bars represent 1$\sigma$ standard deviation.  The vertical dashed line shows the 
occurrence time of the lightning (13:35:55 UT).
%(c) The same as (b), but for monitoring post 6 (black), 5 (red), and 4 (blue).
%(d) The same as (b), but for monitoring post 3 (black), 2 (red), and 1 (blue).
 }
 \label{fig:growth_MPs}
 \end{figure}

\begin{figure}
\noindent\includegraphics[scale=0.4]{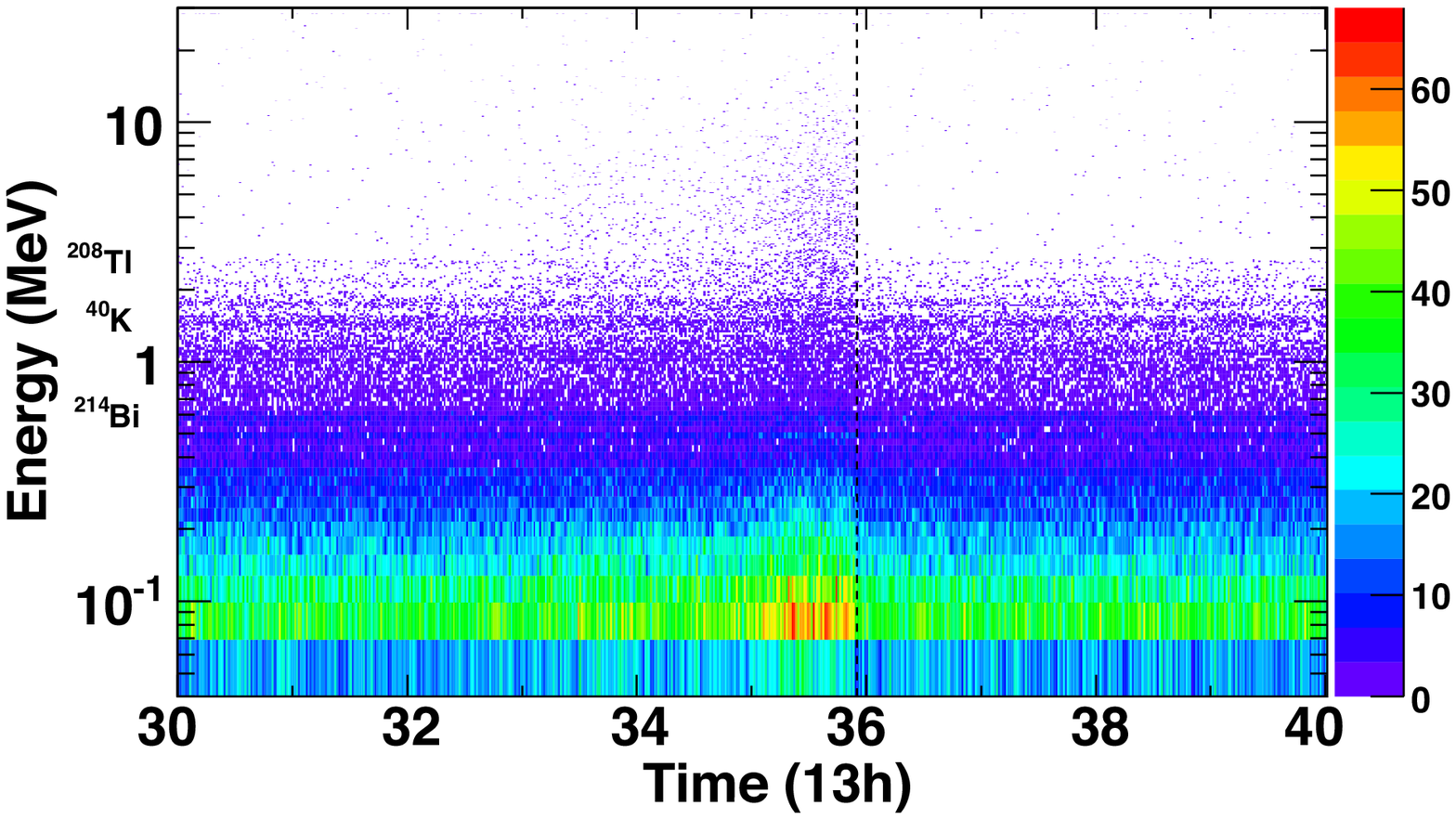}\\
\noindent\includegraphics[scale=0.4]{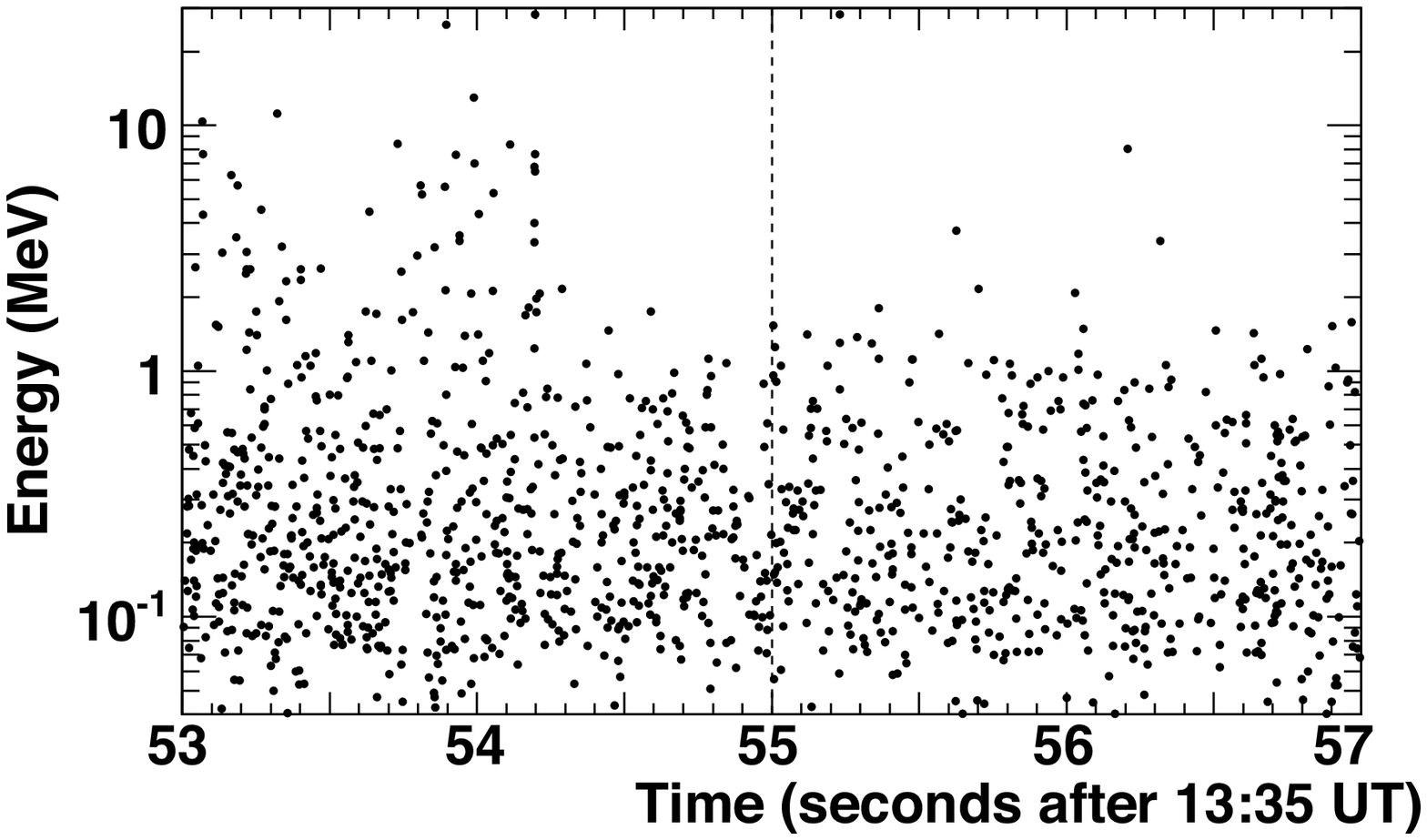}

 \caption{
% (Left) Six sec count histories of 
% the NaI scintillators of daq0 (black), and that of daq1 (red) with anti-coincidence, respectively. 
% Each vertical error bar are statistical 1$\sigma$.
(Top) Photon energies recorded by the NaI scintillator of daq1 with anticoincidence mode during 13:30-13:40 UT. 
A time resolution is 1 sec. Vertical and horizontal axes show the energy in MeV and minutes after 13:00 UT, respectively. 
The vertical dashed line
shows the occurrence time of the lightning, 13:35:55 UT. The color bar denotes counts in each 
bin; as the color approaches red, the number of photons recorded increases. 
(Bottom) Same as that for the left panel, but for the interval of 13:35:53$-$13:35:57 UT, with a time resolution of 0.1 ms.}
 \label{fig:daq1_E_vs_t}
 \end{figure}

Another interesting observation, in addition to the relationship with the lightning flash, was that the measured duration of the present event, 163 s, is 
2$-$4 times longer than that observed by other GROWTH events~\citep{growth_2007,growth_2011}. 
This could be owing to the fact that our detectors, especially daq1, were facing the central region of the 
gamma-ray source for an extended period (Fig.~\ref{fig:image}). 
This favors the detection of $>$10 MeV gamma rays by daq1. 
In this event, the flux of $>$10 MeV gamma rays was the highest energy part of the bremsstrahlung photons
that are emitted by electrons accelerated to a few tens of MeV. 
As known from the bremsstrahlung cross section, such photons, with energies close to  
that of primary electrons, are projected forward in a narrow cone. %This narrow-beaming feature 
%has also been shown by a detailed Monte Carlo simulation~\citep{Dwyer_MC2003}. 
In addition, such high energy photons undergo less Compton scatterings than lower energy photons that tend to be
spread over a wide area. Therefore, it is likely that the extent of the $>$10 MeV gamma rays observed in this event almost equals 
that of the whole acceleration region. 

The 36 s duration of the $>$10 MeV gamma rays detected by daq1 was approximately one fifth of that of the lower energy photons, 163 s.
From the $<$3 MeV observations we found that the extent is at most $\sim$800 m (Fig.~\ref{fig:image});
hence, the extent of $>$10 MeV gamma rays can be inferred as being $\sim$180 m, which is 
almost equal to the acceleration region in the thundercloud.
This is smaller than the region of a positively-charged layer located at the base of 
an electrically active phase of winter thunderclouds~\citep{KM_Obs1994}, which can extend a few kilometers, 
suggesting that only a small part of the electric field is sufficient to give the electrons energies of a few tens of MeV. 

\begin{acknowledgments}
The present work is supported in part by the Suimitomo Foundation, 
the Special Postdoctoral Research Project for Basic Science in RIKEN, and
the Grant-in-Aid for Young Scientists 24740183.
 
\end{acknowledgments}

\clearpage
% ---------------------------------------------------------------------------
 % ---------------------------------------------------------------------------
% ---------------------------------------------------------------------------
\clearpage
%
% 4 Ebands light curves for daq0 and daq1
% ---------------------------------------------------------------------------
 %------------------------------------------------------------------------------
%\begin{figure}
%\noindent\includegraphics[scale=0.7]{Figures/Fig_Hratio_1330-1340_daq01_3-10MeVover0.04-3MeV.eps}
%\noindent\includegraphics[scale=0.7]{Figures/Fig_daq01_6se_HardnessRatio0.04-0.3MeVover3-10MeV_1330-1340.eps}
%\noindent\includegraphics[scale=0.5]{Figures/Fig_daq01_4Ebands_lcs_30sec_1310-1400.eps}
% \caption{Ratios of the 3$-$10 MeV energy band to the 0.04$-$0.3 MeV one for daq0 (black) and daq1 (red), calculated by the
% 6-sec count rates of daq0 and daq1 obtained over 13:30$-$13:40 UT.} 
% \label{fig:daq01_Hratio}
% \end{figure}
 % ---------------------------------------------------------------------------
%\begin{figure}

% \caption{
%}
% \label{fig:daq1_E_vs_t}
% \end{figure}

% \begin{table}
% \caption{}
% \end{table}
%
% ---------------
% TWO-COLUMN figure/table
%
% \begin{figure*}
% \noindent\includegraphics[width=39pc]{samplefigure.eps}
% \caption{Caption text here}
% \end{figure*}
%
% \begin{table*}
% \caption{Caption text here}
% \end{table*}
%
% ---------------
% EXAMPLE TABLE
%
%\begin{table}
%\caption{Time of the Transition Between Phase 1 and Phase 2\tablenotemark{a}}
%\centering
%\begin{tabular}{l c}
%\hline
% Run  & Time (min)  \\
%\hline
%  $l1$  & 260   \\
%  $l2$  & 300   \\
%  $l3$  & 340   \\
%  $h1$  & 270   \\
%  $h2$  & 250   \\
%  $h3$  & 380   \\
%  $r1$  & 370   \\
%  $r2$  & 390   \\
%\hline
%\end{tabular}
%\tablenotetext{a}{Footnote text here.}
%\end{table}
\begin{table}
\caption{30-s count increases in percent and the corresponding statistical significance during 13:34$-$13:36 UT}
\label{tab:sigma}
\begin{tabular}{cccccc}\hline
                & \multicolumn{5}{c}{Detector} \\ \hline
Time\footnotemark[1]       &   daq0           & daq1            & MP9              & MP8      & MP7      \\ 
%13:32:00  & 2.1\% ($1.1\sigma$) &0.5\% ($0.34\sigma$)& 9.4\%($1.4\sigma$)  & 1.1\% ($0.16\sigma$)  &  -1.1\% ($-0.10\sigma$)  \\
%13:32:30  & 0.8\% ($0.46\sigma$) &4.5\% ($2.8\sigma$) & 24\% ($3.4\sigma$) & 2.4\% ($0.35\sigma$) &  1.5\% ($0.23\sigma$)\\
%13:33:00  & 1.4\% ($0.77\sigma$)& 9.3\% ($5.8\sigma$) & 29\% ($3.9\sigma$)& 9.4\%  ($1.4\sigma$) & 5.6\% ($0.83\sigma$)\\
%13:33:30  & 6.1\%  ($3.3\sigma$)& 17\%  ($11\sigma$)& 20\% ($2.8\sigma$)& 5.0\%  ($0.73\sigma$)& 6.8\% ($1.0\sigma$)  \\
13:34:00  & 9.4\%  ($6.8\sigma$)& 17\% ($10\sigma$)& 26\% ($3.5\sigma$)& 15\%  ($2.2\sigma$)  & 15\% ($2.2\sigma$)\\
13:34:30  & 13\%   ($6.8\sigma$)& 21\% ($13\sigma$)& 42\% ($5.6\sigma$)& 41\% ($5.6\sigma$)& 11\% ($1.6\sigma$) \\
13:35:00  & 10\%  ($5.5\sigma$)& 52\% ($30\sigma$) &37\% ($5.1\sigma$)& 62\%  ($8.1\sigma$) & 38\% ($5.2\sigma$)\\
13:35:30  & 14\%  ($7.7\sigma$)& 55\% ($32\sigma$)& 18\% ($2.5\sigma$)& 32\%  ($4.4\sigma$)& 26\% ($3.7\sigma$)\\
%13:36:00  & 3.9\% ($2.1\sigma$)& 2.7\% ($1.8\sigma$)& 0.1\% ($0.014\sigma$)& -1.2\% ($-0.18\sigma$)&1.5\% ($0.23\sigma$)\\
\hline
\end{tabular}

\footnotetext[1]{Start time.}
\end{table}
% See below for how to make landscape/sideways figures or tables.

\end{document}